# Metamaterial model of a time crystal

Igor I. Smolyaninov

*Department of Electrical and Computer Engineering, University of Maryland, College Park, MD 20742, USA*

**Propagation of monochromatic extraordinary light in a hyperbolic metamaterial is identical to propagation of massive particles in a three dimensional effective Minkowski spacetime, in which the role of a timelike variable is played by one of the spatial coordinates. We demonstrate that this analogy may be used to build a metamaterial model of a time crystal, which has been recently suggested by Wilczek and Shapere. It is interesting to note that the effective single-particle energy spectrum in such a model does not contain a static ground state, thus providing a loophole in the proof of time crystal non-existence by P. Bruno.**

In a series of recent papers Wilczek and Shapere considered the possibility that either classical [1] or quantum [2] system may display periodic motion in their lowest-energy state, thus forming a time analogue of crystalline spatial order. This theoretical concept was followed by an experimental proposal by Zhang *et al.* to create a space-time crystal of trapped ions by confining ions in a ring-shaped trapping potential with a static magnetic field [3]. However, soon thereafter Bruno pointed out that Wilczek's rotating soliton is not the correct ground state of the model, and that a static solution with a lower energy can be found [4]. Moreover, Bruno also formulated a more general no-go theorem claiming the impossibility of quantum time crystals [5]. This theorem seems to indicate that unlike ordinary crystals in space, spontaneous symmetry breaking leading to formation of "crystal order" in time is impossible.

In a parallel development, while this very interesting scientific controversy was playing out, it was realized that it is possible to design an electromagnetic hyperbolic metamaterial in such a way that one of its spatial coordinates becomes "timelike" [6]. As a result, this spatial coordinate may play the role of time in an effective "metamaterial spacetime": propagation of monochromatic extraordinary light in such a metamaterial is identical to propagation of massive particles in a three-dimensional (3D) Minkowski spacetime [7]. Moreover, nonlinear optics of such metamaterials appears to be similar to a 3D version of general relativity [8]. It is natural to ask if a metamaterial model of a time crystal may be realized in such a system.

On the one hand, the answer to this question is expected to be trivial "yes": in reality such a crystal would look like a usual crystal in space. On the other hand, careful analysis of the metamaterial system reveals an interesting loophole in the proof of time crystal non-existence by P. Bruno, which makes existence of such a metamaterial version of the time crystal possible. It appears that the effective single-particle energy spectrum in such a model does not contain a static ground state. While this model differs considerably from the system of trapped ions in an external magnetic field considered in [2-4], the detailed analysis of the hyperbolic metamaterial version of the time crystal is interesting because our own 4D Minkowski spacetime may behave as a hyperbolic metamaterial in a very strong magnetic field [9].

Let us start with a brief overview of optical properties of hyperbolic metamaterials. We will consider a non-magnetic uniaxial anisotropic material with (generally frequency-dependent) dielectric permittivities $\varepsilon_x = \varepsilon_y = \varepsilon_1$ and $\varepsilon_z = \varepsilon_2$ in some frequency range around $\omega = \omega_0$. Any electromagnetic field propagating in this material may be expressed as a sum of the "ordinary" and "extraordinary" contributions, each of these being a sum of an arbitrary number of plane waves polarized in the "ordinary" ($\vec{E}$ perpendicular to the optical axis) and "extraordinary" ($\vec{E}$ parallel to the plane defined by the k–vector of the wave and the optical axis) directions. This distinction will be extremely important for the discussion below since it turns out that ordinary and extraordinary photons do not experience the same effective metric. Let us define the extraordinary wave function as $\varphi=E_z$ so that the ordinary portion of the electromagnetic field does not contribute to $\varphi$. Since electromagnetic metamaterials typically exhibit considerable temporal dispersion, we will work in the frequency domain, so that the macroscopic Maxwell equations result in the following wave equation for the spatial distribution of $\varphi_\omega$ [6,7]:

$$-\frac{\omega^2}{c^2}\varphi_\omega = \frac{\partial^2 \varphi_\omega}{\varepsilon_1 \partial z^2} + \frac{1}{\varepsilon_2}\left(\frac{\partial^2 \varphi_\omega}{\partial x^2} + \frac{\partial^2 \varphi_\omega}{\partial y^2}\right) \quad (1)$$

(here we assume that $\varepsilon_1$ and $\varepsilon_2$ are kept constant inside the metamaterial). While in ordinary anisotropic media both $\varepsilon_1$ and $\varepsilon_2$ are positive, in the hyperbolic metamaterials $\varepsilon_1$ and $\varepsilon_2$ have opposite signs [10]. Let us consider the case of $\varepsilon_1 > 0$ and $\varepsilon_2 < 0$. Such a metamaterial may be composed of a metal wire array structure, as shown in Fig.1(a). Under these assumptions equation (1) may be re-written in the form of a 3D Klein-Gordon equation describing a massive scalar field $\varphi_\omega$ in a flat 2+1 dimensional Minkowski spacetime:

$$-\frac{\partial^2 \varphi_\omega}{\varepsilon_1 \partial z^2} + \frac{1}{(-\varepsilon_2)}\left(\frac{\partial^2 \varphi_\omega}{\partial x^2} + \frac{\partial^2 \varphi_\omega}{\partial y^2}\right) = \frac{\omega_0^2}{c^2}\varphi_\omega = \frac{m^{*2}c^2}{\hbar^2}\varphi_\omega \quad (2)$$

in which the spatial coordinate $z=\tau$ behaves as a "timelike" variable, and $m^*=\hbar\omega_0/c^2$ behaves as an effective mass. Note that the components of metamaterial dielectric tensor define the effective metric $g_{ik}$ of this spacetime: $g_{00}=-\varepsilon_1$ and $g_{11}=g_{22}=-\varepsilon_2$. It is easy to check that eq.(2) remains invariant under the "effective Lorentz transformations"

$$z' = \frac{1}{\sqrt{1 - \frac{\varepsilon_1}{(-\varepsilon_2)}\beta}} (z - \beta x) \qquad (3)$$

$$x' = \frac{1}{\sqrt{1 - \frac{\varepsilon_1}{(-\varepsilon_2)}\beta}} \left( x - \beta \frac{\varepsilon_1}{(-\varepsilon_2)} z \right),$$

where $\beta$ is the effective Lorentz boost. The opposite signs of $\varepsilon_2$ and $\varepsilon_1$ ensure that the spatial coordinate z plays the role of time in the Lorentz-like symmetry described by eqs.(3). The extraordinary photon dispersion law

$$\frac{k_z^2}{\varepsilon_1} + \frac{k_x^2 + k_y^2}{\varepsilon_2} = \frac{\omega_0^2}{c^2} = \frac{m^{*2} c^2}{\hbar^2} \qquad (4)$$

shown in Fig.1(b) also exhibits Lorenz-like symmetry, with $k_z$ playing the role of effective energy and the $(k_x, k_y)$ vector playing the role of momentum. However, it is important to note that $k_{xy}=0$ points marked by the green dots in Fig.1(b) are missing in the single-particle spectrum. Indeed, the decomposition of extraordinary field into plane waves does not contain the "static" $k_{xy}=0$ component. Similar to photons and electron neutrinos in vacuum, extraordinary photons cannot be stopped. Extraordinary photons having $k_{xy}=0$ do not exist. Such photons may not have nonzero $E_z$. Absence of such extraordinary photon states provides a loophole in Bruno's proof of time crystal non-existence for the metamaterial Minkowski spacetime.

Let us demonstrate how this loophole may indeed be used to create a metamaterial version of time crystal. Once again, we must emphasize that this task is almost trivial: in reality such a crystal would look just like a usual crystal in space. In fact, such a metamaterial time crystal may be considered as a particular case of recently observed photonic hypercrystals, which result due to spatial modulation of hyperbolic metamaterials induced by various stimuli [11]. However, a metamaterial version of a time crystal may potentially be mapped onto time crystal geometries which may exist in our own 4D Minkowski spacetime. For example, if electron neutrinos have mass, their dispersion law may look similar to Fig.1(b) since $k=0$ points may be prohibited by lepton charge conservation.

An obvious way to create a simple metamaterial time crystal is to use effective gravitational force, which exists between the extraordinary photons in nonlinear hyperbolic metamaterials [8]. Such an effective gravity results from "bending" of the 2+1 dimensional Minkowski spacetime (2) by the nonlinear optical Kerr effect. Indeed, in the presence of nonlinear optical effects the dielectric tensor of the metamaterial (which according to eq.(2) defines its effective metric) may be written as

$$\varepsilon_{ij} = \chi^{(1)}_{ij} + \chi^{(2)}_{ijl} E_l + \chi^{(3)}_{ijlm} E_l E_m + ... \tag{5}$$

It is clear that eq.(5) provides coupling between the matter content (extraordinary photons) and the effective metric of the "metamaterial spacetime". In a centrosymmetric material all the second order nonlinear susceptibilities $\chi^{(2)}_{ijl}$ must be equal to zero. On the other hand, the third order terms may provide gravity-like coupling between the effective metric and the energy-momentum tensor. In the weak gravitational field limit the Einstein equation

$$R^k_i = \frac{8\pi\gamma}{c^4}\left(T^k_i - \frac{1}{2}\delta^k_i T\right) \tag{6}$$

is reduced to

$$R_{00} = \frac{1}{c^2}\Delta\phi = \frac{1}{2}\Delta g_{00} = \frac{8\pi\gamma}{c^4}T_{00} \quad , \tag{7}$$

where $\phi$ is the gravitational potential [12]. Since in the metamaterial spacetime $g_{00}$ is identified with $-\varepsilon_1$, comparison of eqs. (5) and (7) indicates that the third order terms, which are typically associated with the optical Kerr effect do indeed act like gravity. The effective gravitational constant $\gamma^*$ and the third order nonlinear susceptibility $\chi^{(3)}$ of the hyperbolic metamaterial appear to be connected as [8]

$$\frac{4\gamma^*}{c^2\omega_0^2\varepsilon_2} = \chi^{(3)} \qquad (8)$$

As noted in [7], extraordinary light rays in a hyperbolic metamaterial behave as particle world lines in a 2+1 dimensional Minkowski spacetime. Moreover, due to effective gravitational self-interaction they form spatial solitons [8]. Therefore, two extraordinary rays (solitons) launched into the metamaterial at different angles/locations (as shown in Fig.2) will behave as world lines of two particles, which experience mutual weak gravitational attraction. Helical rays formed by two mutually rotating spatial solitons have been observed in [13] and many other similar experiments. Using eq.(8), we may calculate the parameters of such helical rays in a typical hyperbolic metamaterial from the balance of effective gravity and the effective centripetal force. For example, the effective ray "velocity" in this system

$$\vec{v}_{xy} = \frac{d\vec{r}_{xy}}{dz} \qquad (9)$$

may be determined as

$$\vec{v}_{xy}^2 = \gamma^* m^* \qquad (10)$$

The helical solitons form a periodic structure in $z$ direction, which plays the role of time in our model. Thus, they form a metamaterial analogue of a time crystal. Note that such a time crystal is allowed as a direct consequence of the above mentioned loophole in the Bruno's proof of time crystal non-existence. Since $k_{xy}=0$ points in the dispersion law are missing, there is no static non-rotating ground state in the metamaterial system. We should also point out that eq.(8) demonstrates that ordinary rays propagating in *z* direction (thus having $k_{xy}=0$) may not form a spatial soliton via the same effective gravity mechanism, and thus be considered as a true static ground state. Indeed, the effective gravitational constant in eq.(8) depends on $\varepsilon_2=\varepsilon_z$, which cannot be felt by ordinary photons.

Finally, let us address the issue of losses in hyperbolic metamaterials, which may limit validity of our model. Metamaterial losses may indeed attenuate extraordinary soliton intensity, and make our model aperiodic in z direction. However, our consideration was limited to monochromatic extraordinary rays. As demonstrated in ref. [14], a lossless hyperbolic metamaterial design is possible in a narrow frequency range.

In conclusion, we have presented a model of time crystal, which is based on unusual optics of hyperbolic metamaterials. Propagation of monochromatic extraordinary light in a hyperbolic metamaterial is identical to propagation of massive particles in a three dimensional effective Minkowski spacetime, in which the role of a timelike variable is played by one of the spatial coordinates. We have demonstrated that this analogy may be used to build a metamaterial model of a time crystal, which has been recently suggested by Wilczek and Shapere. It is interesting to note that the effective single-particle energy spectrum in such a model does not contain a static ground state, thus providing a loophole in the proof of time crystal non-existence by P. Bruno. The proposed metamaterial version of a time crystal may potentially be mapped onto time crystal geometries which may exist in our own 4D Minkowski spacetime.

**Figure Captions**

**Figure 1.** (a) Schematic views of a wire array hyperbolic metamaterial. The metal wires have negative dielectric constant $\varepsilon_m$, while $\varepsilon_d$ is positive for the dielectric background. Such a geometry may lead to opposite signs of $\varepsilon_z = \varepsilon_2 < 0$ and $\varepsilon_{xy} = \varepsilon_1 > 0$ of the metamaterial. (b) Hyperbolic dispersion relation of extraordinary photons is illustrated as a surface of constant frequency in k-space. Since $z$ coordinate is "timelike", $k_z$ behaves as an effective "energy". Note that $k_{xy}=0$ points marked by the green dots are missing in the single-particle spectrum. If electron neutrinos have nonzero mass, their dispersion law may look similar, since $k=0$ points may be prohibited by lepton charge conservation.

**Figure 2.** Simple metamaterial model of a time crystal composed of helical extraordinary rays formed by two mutually rotating spatial solitons. $Z$ coordinate plays the role of effective time. Since $k_{xy}=0$ points in the dispersion law are missing, there is no static non-rotating ground state in such a system.

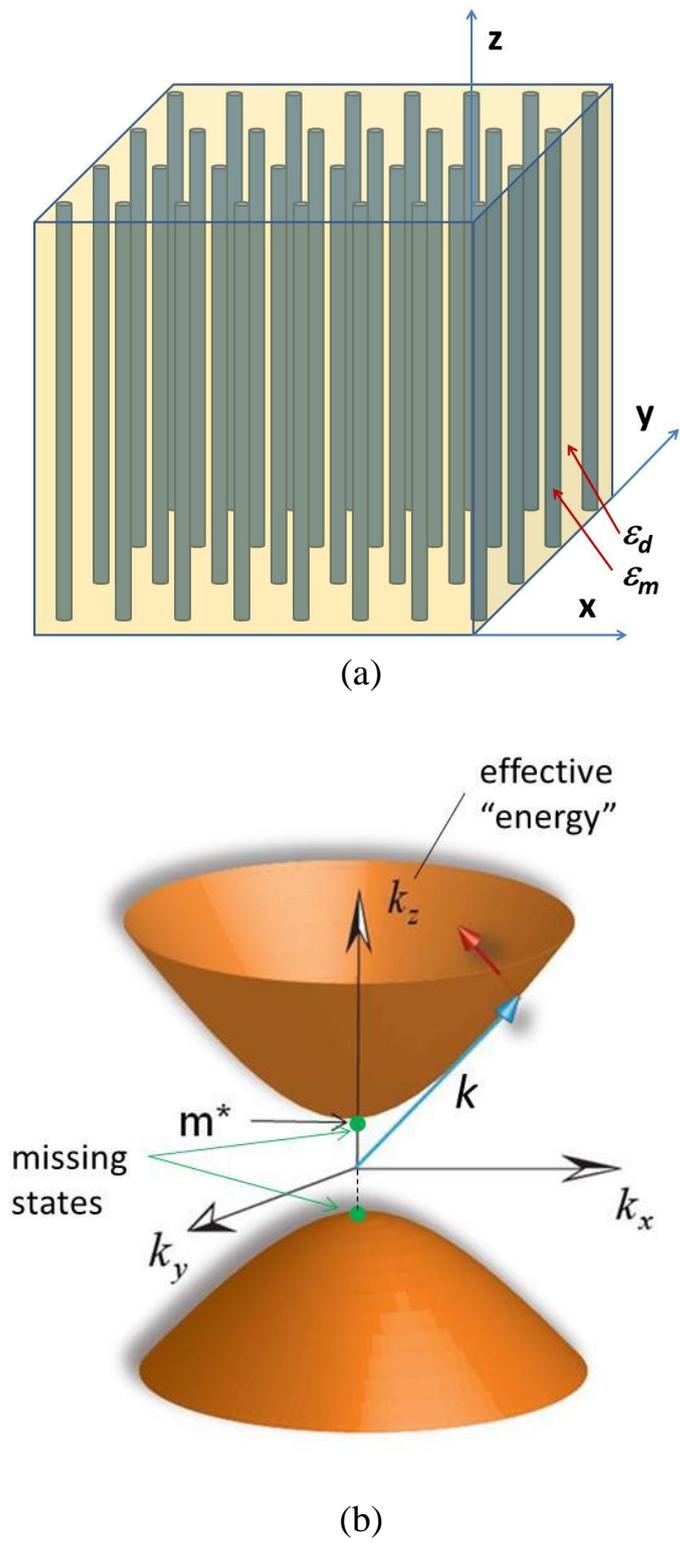

(a)

(b)

Fig. 1

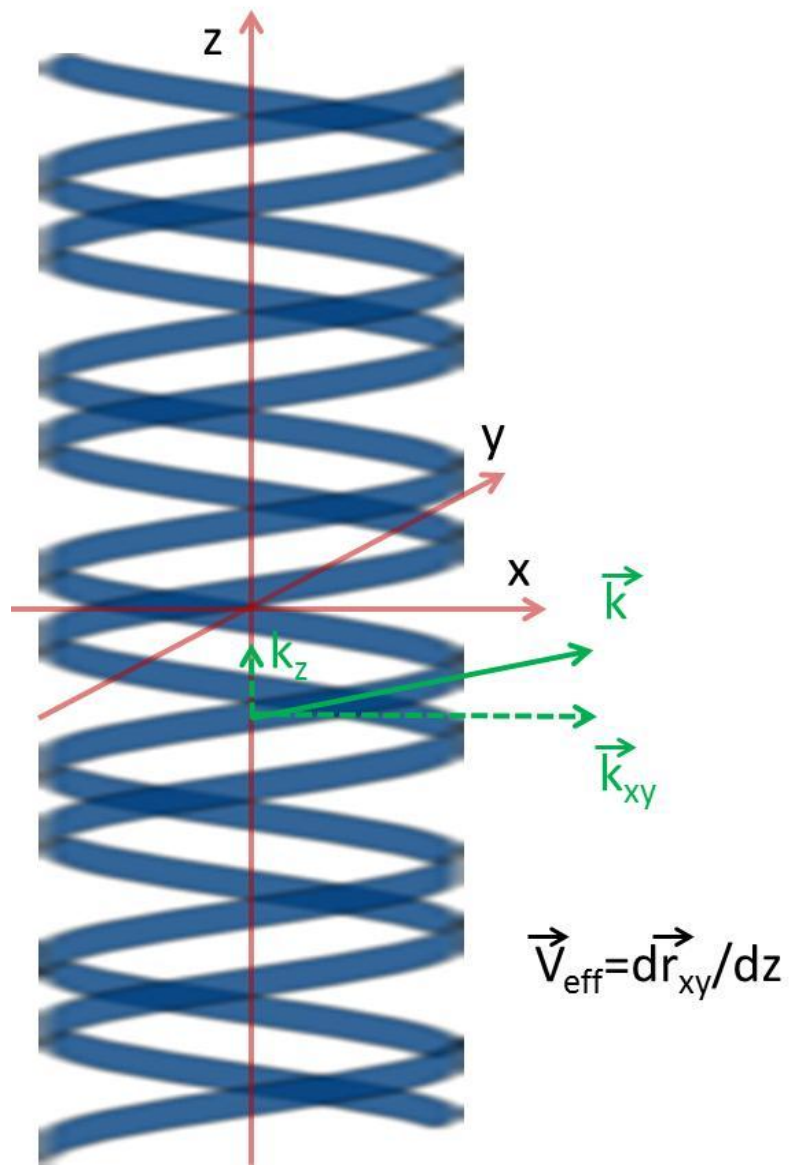

Fig. 2